\documentclass[reprint,aps, pre,showpacs, bibnotes, footinbib]{revtex4-1}
\pdfoutput=1
\usepackage[usenames,dvipsnames]{xcolor}
\usepackage{graphicx}
\usepackage{graphicx,amsfonts}
\usepackage{epsf}
\usepackage{amsmath}
\usepackage{verbatim}

\usepackage{textcomp}
\usepackage{amssymb}

\usepackage{color}
\definecolor{darkblue}{rgb}{0,0,0.5}
\definecolor{darkred}{rgb}{0.5,0,0}
\usepackage[colorlinks=true,urlcolor=darkblue,citecolor=darkblue,linkcolor=darkred,hyperfootnotes=false]{hyperref}

\newcommand{\m}[1]{\mathcal{#1}}
\newcommand{\W}{\mathcal{W}}
\newcommand{\Wd}{\mathcal{W}_{\text{d}}}

\newcommand{\Ha}{\mathcal{H}}

\newcommand{\ps}{p_{\text{s}}}

\newcommand{\dlambda}{\dot{\lambda}}

\newcommand{\B}{\mathcal{B}}

\newcommand{\erfc}{\text{Erfc}}

\begin{document}
\title{Duration of local violations of the second law of thermodynamics along 
single trajectories in phase space}

\author{Reinaldo Garc\'ia-Garc\'ia}
\email{reinaldo.garcia@cab.cnea.gov.ar}
\affiliation{Centro At\'omico Bariloche and Instituto Balseiro, 8400 S. C. de Bariloche, Argentina}
\author{Daniel Dom{\'i}nguez}
\affiliation{Centro At\'omico Bariloche and Instituto Balseiro, 8400 S. C. de Bariloche, Argentina}

\begin{abstract}
We define the {\it violation fraction} $\nu$ as  the cumulative fraction of time that 
the entropy change is negative during single realizations of processes in phase space.
This quantity depends both on the number of degrees of freedom $N$
and the duration of the time interval $\tau$. In the large-$\tau$  and large-$N$ limit  we show that, 
for ergodic and microreversible systems, the mean value of $\nu$ scales as 
$\langle\nu(N,\tau)\rangle\sim\big(\tau N^{\frac{1}{1+\alpha}}\big)^{-1}$. The
exponent $\alpha$ is positive and generally depends on the protocol for 
the external driving forces, being $\alpha=1$ for a constant drive. 
As an example, we study a nontrivial model where the fluctuations
of the entropy production are non-Gaussian: an elastic line driven at a constant rate 
by an anharmonic trap. In this case we show that the scaling of 
$\langle \nu \rangle$ with $N$ and $\tau$ agrees with our result.
Finally, we discuss how this scaling law may break down in the vicinity of a continuous 
phase transition.
\end{abstract}

\maketitle
\section{Introduction}
The energy changes associated to processes occurring in mesoscopic objects exhibit a stochastic
nature due to the effect of fluctuations. This implies that for single trajectories in phase
space, thermodynamic observables like work, heat, and entropy changes are random quantities.
This fact constitutes the basis for stochastic thermodynamics~\cite{Seifert-Review}.
In this context,  the stochastic entropy produced during a certain protocol can 
be negative for certain rare trajectories, a fact that is in  apparent contradiction
with the second law of thermodynamics.
Fluctuation theorems~\cite{Evans-Cohen-Morris,Gallavotti-Cohen,Kurchan,Lebowitz,Jarzynski,Crooks} 
state that those trajectories where the entropy production is negative occur with an exponentially
small weight in comparison to the trajectories where the entropy production is positive
\begin{equation}
 \label{DFT}
\frac{P(S)}{P^{\m{T}}(-S)}=\exp(S),
\end{equation}
where $\m{T}$ represents a transformation like time-reversal, or 
the transformation to a {\it dual} dynamics,
or their composition (see~\cite{Jarzynski1, us1, us2} for a simple definition of the dual dynamics),
and $S$ represents a trajectory-dependent thermodynamic quantity, as 
work, heat, or more generally, different forms of trajectory-dependent entropy production,
with the symmetry $S^\m{T}=-S$.
These relations, initially proved for deterministic and stochastic Markovian
systems, have also been extended to the case of non-Markovian dynamics~\cite{Seifert-non-Markov,
Ohkuma-Ohta,Mai-Dhar,BiroliFTs,Zamponi,Klages,me}.
Furthermore, 
they have been widely tested in experiments~\cite{Wang,Trepagnier,bustamante,carberry,ritort,Gupta}.

Eq.(\ref{DFT}) is commonly denominated a detailed fluctuation theorem (DFT). From it, the relation
$\langle\exp(-S)\rangle=1$, also known as an integral fluctuation theorem (IFT), can be 
obtained. 
From the IFT and Jensen's inequality, the second law of thermodynamics
is obtained as $\langle S\rangle\ge0$, independently of the protocol and the duration of the process. 
Very large systems, where $N\sim10^{23}$, naturally suppress fluctuations since
the mean value of $S$ grows as $N$ for large $N$, while
fluctuations grow as $\sqrt{N}$, a fact that we know from the most elementary
courses of statistical mechanics. In this case $\langle S\rangle\approx S$, 
which means that for any realization
of any process one has $S\ge0$. However, in small 
mesoscopic systems, fluctuations are large and realizations are 
possible such that $S<0$ for not too large time intervals. 
Using some abuse of language (see for 
instance~\cite{Wang}), we can name these  intervals as {\it local violations
of the second law of thermodynamics}, where ``local'' means at a single trajectory level.

We remark that the aforementioned local violations do not represent
true violations of the second law of thermodynamics, which is, perhaps, the most solid law of nature.
The entropy production remains always non-negative in average, however,
it is  interesting to characterize the statistics of the total time of occurrence
of these events. In particular, if we consider a time interval $[0,\tau]$, we will be interested
in calculating the total fraction of time $\nu$ where the stochastic entropy production is negative, which we 
name  as the {\it violation fraction}.

\begin{figure}
\begin{center}
  \includegraphics[scale=0.8]{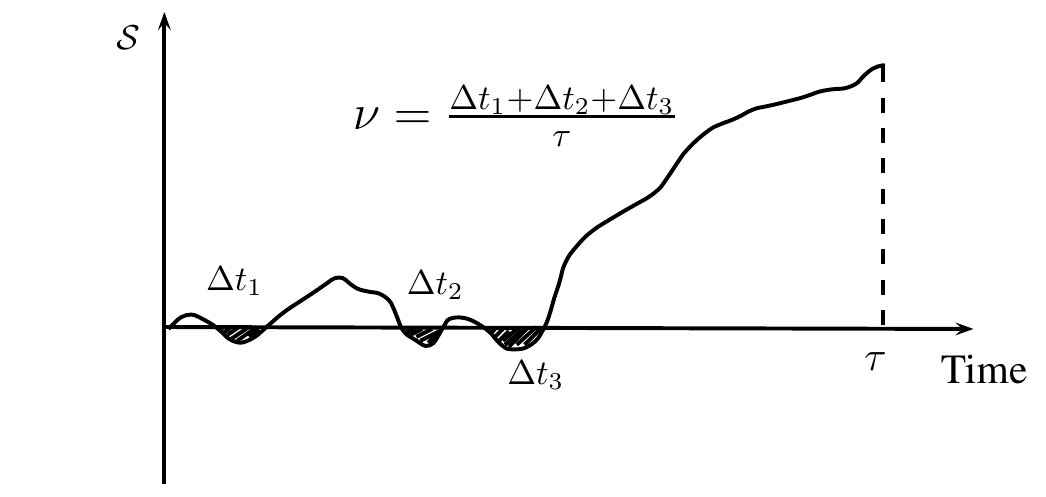}
  \caption{A sketch of the evolution of the stochastic entropy change for a
  particular trajectory in phase space of a mesoscopic system. The violation fraction
  $\nu$ is defined as the relative duration of the violation sectors (where $S<0$) with respect to the
  total time $\tau$.}
  \label{fig1}
  \end{center}
\end{figure}

In Fig. \ref{fig1} we sketch the evolution of the
entropy change as a function of time for a stochastic trajectory of a small system. The shadowed
regions are the violation sectors (where $S<0$)  
and the cumulative duration of these regions divided by the total
time $\tau$ defines the  fraction $\nu$.

This study is important in those cases where one only needs to consider a small number of realizations of a given
process at the mesoscopic level. The entropy production is a self-averaging quantity,
which implies that for single realizations in macroscopic systems
one is already observing the averaged entropy production. That is why, in spite of the fact that the second
law of thermodynamics forbids the spontaneous emergence of order from disorder 
in isolated systems {\em only in average}, we meet this limitation {\em at every single realization} of the given process.
This is clearly not the picture at the mesoscopic scale, where this constraint is relaxed
at single trajectories, and order can sometimes emerge from disorder.

Additionally, our theory may be relevant in some other practical situations.
For example, it is known that the efficiency of any thermal machine is bounded
by virtue of the second law of thermodynamics. However, if one optimizes the external
protocols and accomodates the values of 
$\tau$ and $N$ such that the average violation fraction exhibits its larger possible
value, our motor will operate in a scenario where the occurrence of entropy-consuming trajectories is especially
propitious, a fact that could enhance the efficiency of the machine.
This possibility will be considered in more detail in future works.

Before proceeding, we would like to advance our main result. For the generic kind of systems we will
consider in this work,
the mean violation fraction exhibits, at large times, the simple behavior 
$\langle\nu(N,\tau)\rangle\sim A(N)/\tau$,
where $A(N)$ is a model-dependent function of $N$ which vanishes in the thermodynamic limit 
as $A(N)\sim A_\infty/N^\frac{1}{1+\alpha}$, with $A_\infty$ a non-universal, model-dependent constant,
and $\alpha>0$, an exponent which depends on the particular form of the protocol.

The rest of the paper is organized as follows. In the next section we pose our problem 
and introduce the main concepts we need to tackle it. In section \ref{sec:main}, we prove our main result. 
Later, we study the illustrative case of a linear chain of harmonic oscillators driven
at a constant rate by an anharmonic trap in section \ref{sec:chain}. We briefly
discuss how the strong fluctuations around a continuous phase transition 
may lead to a failure of the derived scaling law in section \ref{sec:phase-transition}.
Finally, we present some conclusions and future projections of this work in section \ref{sec:conclusions}.

\section{Generalities}
\label{sec:gen}
In order to characterize the local violations of the second law of thermodynamic
associated to generic processes occurring in mesoscopic systems,
we have to study the zero-crossings properties 
of the stochastic entropy production $S(t)$. An important quantity is the residence time,
which is defined as the amount of time, up to time $\tau$, 
that a given stochastic process $X(t)$ 
spends in the semi-space  $X(t)<0$ (or $X(t)>0$). 
The probability density function (PDF) for this quantity
is closely related to the persistence, e.g. 
the probability that the process does not cross the line $X=0$ up 
to time $\tau$~\cite{Dornic}, which has 
been studied in many different contexts. 
For example, in~\cite{Dornic} it has 
been used to characterize the ordering dynamics of the Ising model 
(where the process $X(t)$ is the global or local magnetization),
while in~\cite{Newman} it has been used to characterize the zero-crossings
properties of a diffusing field starting from random initial conditions
($X(t)$ is the value of the diffusing field at a fixed point in space, as
a function of time).

Generally speaking, the computation of the full PDF of the residence time 
for a given stochastic process is a difficult task, even when the process
is Markovian. In our case, we choose $X$ to be a stochastic entropy
production (see definitions below) which is generally non-Markovian even
if the system performs Markov dynamics. 
Since an exact determination of the full PDF for arbitrary
protocols and energy landscapes is out of order, we focus in deriving some
general results for the first moment of this PDF, the average violation fraction.
This means that the results presented here are expected to be robust upon
variations in the Hamiltonian of the system, as long as these changes do not
affect the main assumptions we use in our derivation, which we now discuss.

We consider ergodic, microreversible, 
and extensive systems driven by a set of external, 
well controled parameters
(which we denote by $\lambda(t)$).
We point out that ergodicity and microreversibility (the latter for the closed system composed by the 
system under study and its environment) are minimal conditions for fluctuation relations to hold.
On the other hand, in order to guarantee the extensivity of the thermodynamic observables,
it is reasonable to assume that the interactions among the different particles in the system
are short-ranged.

For constant protocols, the system is able to reach a unique steady-state
characterized by the PDF $\ps(x;\lambda)=\exp[-\phi(x;\lambda)]$, 
where $x$ denotes a configuration in the phase space. $x$ may be
a single variable, a vector, or a field as well.
We discuss equilibrium and non-equilibrium steady-states (NESS) on the same footing. Furthermore,
we assume that the system is initially prepared in the steady-state corresponding 
to $\lambda(0)=\lambda_0$,
and for simplicity we assume that the system is connected to a single thermal bath. A more general
analysis, considering several reservoirs, will be published elsewhere.
We introduce the process $S(t)$ as follows
\begin{equation}
 \label{proc-entropy}
S(t)=\int_0^tdt'\dot{\lambda}(t')\partial_\lambda\phi(x(t');\lambda(t')),
\end{equation}
which depends on the particular trajectory of the system in the phase-space during its evolution.
The time instant $t$ should be considered as an arbitrary point inside the interval $[0,\tau]$,
where the violation fraction will be measured. We also remark that $\tau$ can be choosen arbitrarily.

For equilibrium steady-states, $S(t)$ is the total entropy change, proportional
to the dissipated work~\cite{Crooks} $\beta\Wd=\beta\W-\beta\Delta F$, where 
$\W$ is the thermodynamic work introduced by
Jarzynski~\cite{Jarzynski}, whereas $F$ is the free energy and $\beta$ the inverse temperature.
To see this, note that for a system with Hamiltonian $\m{H}(x;\lambda)$, the thermodynamic
work reads~\cite{Jarzynski}
\begin{equation}
 \label{work-def-gen}
 \m{W}(t)=\int_0^tdt'\dot{\lambda}(t')\partial_\lambda\m{H}\big(x(t');\lambda(t')\big).
\end{equation}
On the other hand, we have that $\phi(x;\lambda)=\beta\m{H}(x;\lambda)-\beta  F(\beta,\lambda)$, thus,
eq. (\ref{proc-entropy}) reads in this case:
\begin{align}
 \label{work-entropy-rel}
S &=\beta\int_0^tdt'\dot{\lambda}(t')\partial_\lambda
\big[\m{H}\big(x(t');\lambda(t')\big)-F\big(\beta,\lambda(t')\big)\big]\nonumber\\
 &\equiv\beta\m{W}-\beta\Delta F
\end{align}
For NESS, $S(t)$
corresponds to the dissipation function of Hatano and Sasa~\cite{Hatano-Sasa}, 
which is the correct choice in order to extend the second law of thermodynamics to NESS.

Having clarified the context and the main assumptions we will need in what follows, we 
proceed now to derive our main result in the next section.
\section{Scaling law for the mean violation fraction}
\label{sec:main}
Let us consider an ergodic, extensive, and microreversible system initially prepared in its steady state
corresponding to a value of the control parameters $\lambda(0)=\lambda_0$.
We should note first that these conditions imply that the system satisfies 
the IFT (Jarzynski~\cite{Jarzynski} for equilibrium states
or Hatano-Sasa~\cite{Hatano-Sasa} for NESS) at any time $t$:
\begin{equation}
 \label{IFT}
\langle\exp(-S)\rangle=\int dS P(S,t)\exp(-S)=1.
\end{equation}
From eq. (\ref{IFT}) and Jensen's inequality,  $-\ln x\ge1-x$ for $x>0$, we see that
the second law of thermodynamics holds in the strict sense, $\langle S\rangle\ge0$. On the other
hand, the mean entropy production rate, $\langle\dot{S}(t)\rangle$, should also
be non-negative. Thus, $\langle S(t)\rangle$ is non-decreasing by virtue of the second law of thermodynamics.

Let us formally introduce the violation fraction as
\begin{equation}
 \label{viol-def}
 \nu(\tau,N)\stackrel{def}{=}\frac{1}{\tau}\int_0^\tau dt \Theta(-S(t,N)),
\end{equation}
where $\Theta(\bullet)$ is the Heaviside step function. Then, the mean value of this
fraction is given by
\begin{equation}
 \label{mean-viol-frac}
 \langle\nu(\tau,N)\rangle=\frac{1}{\tau}\int_0^\tau dtR(t,N),
\end{equation}
where $R(t,N)=\text{Prob}[S(t,N)<0]$. Quite generally, we may write
$R(t,N)=(1/2)\exp[-\Phi(t,N)]$, with $\Phi(t,N)$ a non-decreasing function
of time such that $\Phi(0,N)=0$. 

Note that $\Phi$ should be non-decreasing
since in the thermodynamic limit
the entropy production per particle converges to its mean in density: $s(t,N)=S(t,N)/N\rightarrow\mu_S(t)=
\langle S(t,N)\rangle/N$,
or equivalently,
$NP(Ns,t,N)\rightarrow\delta(\mu_S(t)-s)$. Given that $\mu_S\ge0$ and $\dot{\mu}_S\ge0$ 
by virtue of the second law of thermodynamics, the probability to find negative values
of $S$ must decrease with time. On the other hand, this function satisfies $\Phi(0,N)=0$
since, given the definition of the entropy production, eq. (\ref{proc-entropy}), we have
that $P(S,0,N)=\delta(S)$, so that $R(0,N)=\int_{-\infty}^0\delta(S)dS\equiv1/2$.
An important point the reader should note is that, if $\int_0^\infty R(t,N)dt<\infty$,
the average violation fraction vanishes for $\tau\rightarrow\infty$ as
\begin{equation}
 \label{viol-time}
 \langle\nu(\tau,N)\rangle=\frac{A(N)}{\tau},
\end{equation}
where
\begin{equation}
 \label{A-n}
 A(N)=\int_0^\infty R(t,N)dt=\frac{1}{2}\int_0^\infty\exp[-\Phi(t,N)]dt.
\end{equation}
Typical non-critical systems with short-ranged interactions
are characterized by relatively small fluctuations in the thermodynamic limit. 
Note that for any stochastic realization of the process $S(t)$ we can write
$S(t)=\langle S(t)\rangle+\delta S(t)$, with $\delta S=S-\langle S\rangle$.
The previous statement, formally written, expresses a fact we have already
discussed in the introduction, say, $\langle S\rangle\sim N$, while $\sqrt{2\langle\delta S^2\rangle}\sim\sqrt{N}$.
Given that $\langle S\rangle$ is non-decreasing as a function of time,
there is a finite time-scale $\tau_c(N)$ so that for times $\tau>\tau_c$ the local violations
of the second law of thermodynamics are not likely to occur. This time-scale can be roughly
determined by the relation $\langle S(\tau_c,N)\rangle=\sqrt{2\langle\delta S^2(\tau_c,N)\rangle}$.
Then, the cumulative violation time $\int_0^\infty R(t,N)dt\propto\tau_c(N)<\infty$.

It is, however, important to note that critical systems or systems where the interactions are long-ranged 
may lead to infinite values
of the given integral since fluctuations are strong in those cases. This fact may induce non-trivial
power laws for the relaxation of the average violation fraction at large times.

We proceed now by noting that at any finite time instant $t\neq0$, one should
have:
\begin{equation}
 \label{lim-phi}
 \lim_{N\rightarrow\infty}\Phi(t,N)=\infty.
\end{equation}
This property is very intuitive. In the thermodynamic limit the entropy production converges
in density to its positive mean, as we have said before. Then, every process ocurring in
any macroscopic system produce positive entropy , which implies that the probability of
the occurrence of local violations of the second law vanishes in the thermodynamic limit.

As a consecuence of eq. (\ref{lim-phi}), for large $N$ the integral in eq. (\ref{A-n}) is dominated by the expansion of $\Phi$
around $t=0$, where the absolute minimum, $\Phi(0,N)=0$, is reached. In words, in the thermodynamic limit
the scaling with $N$ of the mean violation fraction is determined by the behavior of $\Phi$ at short times.
Our arguments hold generally,
given that the correlation vo\-lume
of the system is finite, i.e., that the system is far from a critical point. If it were not the case 
any finite $N$ should be regarded as small irrespectively of how large it is. In that case, the asymptotics of 
$\nu$ will no longer be dominated by the short times, a fact that we discuss in section \ref{sec:phase-transition}.

Before proceeding, we discuss the illustrative case where the PDF of the entropy production
is Gaussian. In this context we have
\begin{equation}
 \label{generic-Gaussian}
P(S,t,N)=\frac{1}{\sqrt{2\pi\Sigma_{S}(t,N)}}\exp\bigg[-\frac{\big(S-m_{S}(t,N)\big)^2}{2\Sigma_{S}(t,N)}\bigg],
\end{equation}
with $\Sigma_S=\langle\delta S^2\rangle$, and $m_S=\langle S\rangle$. An important
identity that comes out within this framework is $\Sigma_S=2m_S$, which can be easily seen by substituing
eq. (\ref{generic-Gaussian}) in (\ref{IFT}).
Then we may write for $R(t,N)$:
\begin{equation}
 \label{R-2}
 R(t,N)=\int_{-\infty}^0P(S,t,N)dS=\frac{1}{2}\erfc\bigg(\frac{1}{2}\sqrt{m_S(t,N)}\bigg), 
\end{equation}
which allows us to identify the function $\Phi(t,N)$ exactly:
\begin{equation}
 \label{R-3}
 \Phi(t,N)=-\ln\erfc\bigg(\frac{1}{2}\sqrt{m_S(t,N)}\bigg).
\end{equation}
In this case we have that $\Phi(t,N)$ depends on $t$ and $N$ implicitly via $m_S(t,N)$. In particular, given
that $m_S(0,N)=0$, we have
\begin{equation}
 \label{phi-exp-m}
 \Phi(t,N)=\sqrt{\frac{m_S(t,N)}{\pi}}+O(m_S(t,N)),
\end{equation}
around $t=0$. Inspired by this result for the Gaussian case, we will assume that at least at short times
one can generally write for arbitrary systems:
\begin{equation}
 \label{guess}
 \Phi(t,N)\approx\Psi\big(m_S(t,N)\big),
\end{equation}
with $\Psi(0)=0$. This assumption is at least plausible, since $m_S$ and $\Phi$ are non-decreasing
functions of time which vanish at $t=0$. In fact, given the relation between the monotonicity of both
functions, one can write exactly $\Phi(t,N)=\Psi\big(m_S(t,N),N\big)$. Then, our assumption means that
we consider the explicit dependence on $N$ as subleading.

Our ansatz for $\Phi$ at short times, eq. (\ref{guess}), is simply an intuitive guess. We do not have
a formal proof for it. Given that this ansatz is exact for Gaussian systems, the results we are about
to derive hold exactly in that case. However, it is not evident at a first glance that this guess also
provide good results when the fluctuations of the entropy production are non-Gaussian. Although we do not
expect it to be considered as a formal proof, in section \ref{sec:chain} we discuss a nontrivial model
with non-Gaussian fluctuations, obtaining a very good agreement with our predictions.

Now, we follow by noting that
$(d/dt)m_S(t,N)|_{t=0}=0$. This 
relation can be obtained by taking the derivative w.r.t. $t$
in both sides of eq. (\ref{proc-entropy}), taking the average, and considering that the system, at $t=0$, is prepared
in its steady-state:
\begin{align}
 \label{proof-rate}
 \frac{d}{dt}m_S(t,N)\bigg|_{t=0} &=\dot{\lambda}(0)\langle\partial_{\lambda_0}\phi(x(0);\lambda_0)\rangle\nonumber\\
 &=\dot{\lambda}(0)\int \partial_{\lambda_0}\phi(x;\lambda_0)\ps(x;\lambda_0)dx\nonumber\\
 &=\dot{\lambda}(0)\int \partial_{\lambda_0}\phi(x;\lambda_0)\exp[-\phi(x;\lambda_0)]dx\nonumber\\
 &=-\dot{\lambda}(0)\partial_{\lambda_0}\int \exp[-\phi(x;\lambda_0)]dx\nonumber\\
 &=-\dot{\lambda}(0)\partial_{\lambda_0}\int \ps(x;\lambda_0)dx\nonumber\\
 &\equiv0.
\end{align}
The previous result implies that at short times the entropy production can be written, to the first non-vanishing
order, as $m_S(t,N)=N\theta t^{1+\alpha}$, with $\alpha>0$, and $\theta>0$ by virtue of the second law of thermodynamics.
On the other hand, we made explicit the fact that $m_S$ is extensive by writing a factor $N$. With this, our ansatz
writes at short times:
\begin{equation}
 \label{guess-1}
 \Phi(t,N)\approx\Psi\big(Nt^{1+\alpha}\big),
\end{equation}

Although we will prove it formally in section \ref{sec:phase-transition}, we would like to point out
that in the present framework one expects that for a constant drive $\alpha=1$. We can appeal to our intuition to understand
this statement. Note that $P(S,t=0,N)=\delta(S)$. Then, one may think that for $t\gtrsim0$ 
one can approximate $P(S,t,N)$ by a Gaussian distribution, since $m_S$ is small and $P(S,t\sim0,N)$ is very
narrowed around $m_S$.
With this in mind, one has that $m_S\approx(1/2)\Sigma_S$ at short times, a consecuence of the fluctuation
theorem eq. (\ref{IFT}). Note that, by definition, $\Sigma_S$ involves 
a doble integral w.r.t.
time where the term $\dot{\lambda}^2=\text{const}$ is multiplied by a correlation function which is nonzero when evaluated
for equal time arguments. Thus, $\Sigma_S$ can be written at short times as $\Sigma_S(t)\approx Bt^2$, being $B$ a constant,
which means that $m_S\sim t^2$, and $\alpha=1$.
The formal proof we give in section \ref{sec:phase-transition} uses the generalized fluctuation-dissipation relation.

The Gaussian approximation at very short times must be valid for large $N$. The non-Gaussian
structure of $P(S,t,N)$ takes some time to show up and for large $N$, given the central limit
theorem, one expects this time to be large since the Gaussian fluctuations around the
mean dominate, while the large deviations from the mean, which are not well described by the central limit
theorem, are extremely rare. An interesting point to be investigated is
to what extent the Gaussian approximation around $m_S$ is enough to describe the full behavior
of $\langle\nu\rangle$ at any time, given that $N$ is large. It is also interesting to investigate
how large should $N$ be in that case.

Let us now continue with our derivation. Substituing (\ref{guess-1}) in (\ref{A-n}), we obtain
\begin{equation}
 \label{A-n-1}
 A(N)\sim\frac{1}{2}\int_0^\infty\exp\big[-\Psi\big(Nt^{1+\alpha}\big)\big]dt=\frac{\varrho(\alpha)}{N^{\frac{1}{1+\alpha}}},
\end{equation}
with
\begin{equation}
 \label{varro}
 \varrho(\alpha)=\frac{1}{2(1+\alpha)}\int_0^\infty\exp\big[-\Psi(u)\big]u^{-\frac{\alpha}{1+\alpha}}du.
\end{equation}
Note that we have introduced the change of variables $u=Nt^{1+\alpha}$ for the evaluation of the integral in
eq. (\ref{A-n-1}).
Now, putting all together, we can write
\begin{equation}
 \label{mean-viol-final-fin}
 \langle\nu(\tau,N)\rangle\sim\frac{A_\infty}{\tau N^\frac{1}{1+\alpha}},
\end{equation}
as we previously announced, with $A_\infty\propto\varrho(\alpha)$. We proceed now
to test our general result by means of the numerical study of a nontrivial model system.

\section{An elastic line dragged by an anharmonic trap}
\label{sec:chain}

We now proceed to test our results by studying a simple, yet illustrative system.
As discussed in the previous section, our main result, eq. (\ref{mean-viol-final-fin}),
holds exactly for systems where the stochastic entropy production is a Gaussian process. For that reason, we must consider
a non-Gaussian model system in order to validate the generality of our approach.
When the entropy production is Gaussian, one can easily consider the corrections to (\ref{mean-viol-final-fin}) up to
any arbitrary order for finite $\tau$ and $N$, since the function $\Phi(t,N)$ is known exactly, as given by
eq. (\ref{R-3}). For generic non-Gaussian models, as the one considered in this section, $\Phi(t,N)$ is hard to determine
analytically, so it is extremely hard to  compute finite-time and finite-$N$ corrections to (\ref{mean-viol-final-fin}).
For that reason, we focus in this section only on the validation of our main result, 
since the determination of the referred corrections  does not provide, in our opinion, relevant and clear enough information
as to pay the cost of performing the tedious calculations involved. 

We study a discrete line consisting of
particles interacting via a short-ranged elastic potential
and coupled to an anharmonic trap centered at $\lambda$. 
We will consider as the external protocol the control of $\lambda(t)$.
The Hamiltonian of this system is given by
\begin{equation}
 \label{syst-hamiltonian}
\Ha=\sum_{i=1}^N(u_{i+1}-u_i)^2+
\frac{k_0}{2}\sum_{i=1}^N(u_i-\lambda)^2+\frac{k_1}{4}\sum_{i=1}^N(u_i-\lambda)^4,
\end{equation}
where $u_i$ are the elastic displacements of the particles, and 
$k_0$ and $k_1$ are positive constants denoting the strength of the parabolic and quartic couplings to
the nonlinear spring (relative to the elastic constant of the chain). 
We assume periodic boundary conditions (PBC).

Given that this system exhibits equilibrium steady states, the entropy production is in this case given
by eq. (\ref{work-entropy-rel}). On the other hand, it is simple to see that for the present system the free energy
does not depend on $\lambda$, and thus one has $S=\beta\W$. For the work, 
we can write
\begin{equation}
 \label{work.def}
\W=\sum_i\int_0^t dt'\dlambda(t')\big(k_0\phi_i(t')+k_1\phi_i^3(t')\big),
\end{equation}
with $\phi_i(t)=\lambda(t)-u_i(t)$,
 a result which follows from eqs. (\ref{work-def-gen}) and (\ref{syst-hamiltonian}).
 We make the reader note that the work, as given by eq. (\ref{work.def}), is a non-Gaussian process. On the other hand,
 we point out that, assuming simple relaxational dynamics for our model system, and performing the
 shift $\phi_i(t)=\lambda(t)-u_i(t)$ in the dynamical equations, we obtain a time-dependent Ginzburg-Landau
 equation (TDGLE) for a magnet in the paramagnetic phase ($T>T_c$) in presence of a time dependent magnetic field:
 \begin{equation}
  \label{TDGLE}
  \dot{\phi}_i(t)=\nabla_i^2\phi_i(t)-k_0\phi_i(t)-k_1\phi_i^3(t)+\dlambda(t)+\xi_i(t),
 \end{equation}
where $\nabla_i^2\phi_i(t)=\phi_{i+1}(t)+\phi_{i-1}(t)-2\phi_{i}(t)$, is the discrete Laplacian, and
$\xi_i(t)$ is a thermal noise with zero mean and variance $\langle\xi_i(t)\xi_j(t')\rangle=2T\delta_{ij}\delta(t-t')$.

\begin{figure}
 \begin{center}
  \includegraphics[width=8.5cm]{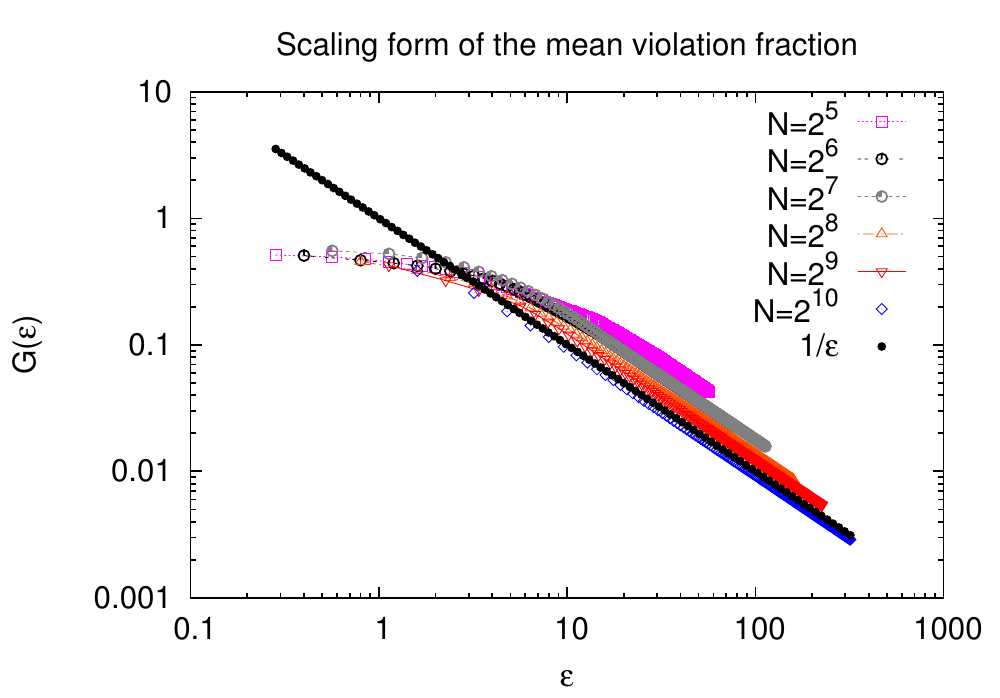}
  \caption{(Color online) Scaling form of $\langle\nu(\tau,N)\rangle$ in the elastic system
  as a function of $\varepsilon=\tau\sqrt{N}$.
For large $N$, the law given by eq. (\ref{mean-viol-final-fin}) with $\alpha=1$ is verified.}
  \label{fig2}
  \end{center}
\end{figure}

We consider a constant drive, in which case $\alpha=1$ is expected in eq. (\ref{mean-viol-final-fin}). 
Taking $\lambda(t)=bt$,
eq. (\ref{TDGLE}) reduces to a TDGLE in presence of a constant magnetic field $b$.
Starting from a flat configuration, $\phi_i=0$ for each $i$, we allow the line to evolve
until it thermalizes with the heat bath, keeping $\lambda=0$ fixed. After that, we start to
move the trap at a constant speed, which means that we suddenly turn on $b$ in (\ref{TDGLE}).
The violation fraction is then periodically sampled as a function of time.

If we introduce the scaling variable $\varepsilon=\tau\sqrt{N}$, we should have, for large $\tau$ and $N$, that 
$\langle\nu(\tau,N)\rangle=G(\varepsilon)\sim\varepsilon^{-1}$.
In Fig. \ref{fig2} we plot the numerical solution of eq. (\ref{TDGLE}) for different system sizes. It can be
seen that for large $N$ we obtain the expected scaling form for the violation fraction. In particular, the
curve with $N=2^{10}$ particles fits very well to our theoretical line at large times. 

\section{Failure around a continuous phase transition}
\label{sec:phase-transition}
We now discuss that the obtained scaling law may fail in the vicinity of a critical point. 
First, let us note that for any protocol
 $\lambda(t)=\lambda_0+\delta\lambda(t)$ we can write at short times 
\begin{equation}
 \label{exp-1}
\langle\partial_\lambda\phi(x(t),\lambda(t))\rangle\approx
\langle\partial_\lambda\phi(x(t),\lambda_0)\rangle+\delta\lambda(t)
\langle\partial^2_\lambda\phi(x(t),\lambda_0)\rangle_{\text{ss}},
\end{equation}
where the double-s subscript means
that the average is taken in the steady-state corresponding to $\lambda=\lambda_0$.
On the other hand, to the same order, the normalization condition 
for the steady-state PDF leads to the relation
\begin{equation}
 \label{corr-0}
 \langle\partial^2_\lambda\phi(x(t),\lambda_0)\rangle_{\text{ss}}=C_r(0),
\end{equation}
where the correlation function $C_r$ is given by
\begin{equation}
 \label{corr}
C_r(t-s)=\langle\partial_\lambda\phi(x(t),\lambda_0)\partial_\lambda\phi(x(s),\lambda_0)\rangle_{\text{ss}}.
\end{equation}
Indeed, we have that the stationary PDF satisfies, for each $\lambda$, the normalization condition
\begin{equation}
 \label{C-norm}
\int\exp[-\phi(x;\lambda)]dx=1.
\end{equation}
Additionally, up to the second order in $\delta\lambda=\lambda-\lambda_0$, we have
\begin{widetext}
\begin{equation}
 \label{C-norm-1}
\exp[-\phi(x;\lambda)]=\exp[-\phi(x;\lambda_0)]
\bigg(1-\delta\lambda\partial_\lambda\phi(x;\lambda_0)
+\frac{1}{2}\delta\lambda^2\bigg[\partial_\lambda\phi(x;\lambda_0)\bigg]^2-
\frac{1}{2}\delta\lambda^2\partial_\lambda^2\phi(x;\lambda_0)+O(\delta\lambda^3)\bigg).
\end{equation}
\end{widetext}
Substituing (\ref{C-norm-1}) in (\ref{C-norm}), and using eq. (\ref{proof-rate}) in order
to drop the linear term in $\delta\lambda$, we immediately obtain (\ref{corr-0}).

We now continue by noting that, using the generalized fluctuation-dissipation relation (see, for instance,
~\cite{Parrondo}), one can write 
\begin{equation}
 \label{FDT}
\langle\partial_\lambda\phi(x(t),\lambda_0)\rangle=\int_0^t ds\chi(t-s)\delta\lambda(s),
\end{equation}
with $\chi(t-s)=\partial_tC_r(t-s)$.
With all this, using (\ref{proc-entropy}), we have at short times
\begin{equation}
 \label{dot-mean}
\dot{m}_S(t)\approx\delta\dot{\lambda}(t)\bigg[\int_0^t ds\chi(t-s)\delta\lambda(s)+\delta\lambda(t)C_r(0)\bigg].
\end{equation}

At this point, we would like to show formally that for a constant drive one must expect $\alpha=1$ in
eq. (\ref{mean-viol-final-fin}). In that case one has
$\delta\lambda(t)=bt$ and we get
\begin{equation}
 \label{dot-mean-const}
\dot{m}_S(t)\approx b^2\bigg[\int_0^t ds\chi(t-s)s+tC_r(0)\bigg],
\end{equation}
leading to the result $\ddot{m}_S(0)=b^2C_r(0)$. Thus, at short times $m_S(t)\approx(1/2)C_r(0)b^2t^2$,
which means that $\alpha=1$, as we claimed before. On the other hand, far from a critical point, one has that
$C_r(0)\sim N$, and the entropy production is extensive, as expected.

We continue now by assuming that our system has linear size $L=N^{1/d}$, with $L$ large but finite,
and that we are in the vicinity of a phase transition. This could be achieved, for example,
by tuning the temperature of the thermal bath to be equal to the critical temperature of the thermodynamic
system, $T_c(\infty)$, (for which $N\sim10^{23}$). The finite system does not exhibit critical behavior, but given that
$T=T_c(\infty)$, the system starts to be critical as we increase the value of $N$.

To continue, let us also assume that the correlation and response
functions are associated to an order parameter.
The scaling form of the response function depends, for an infnite system,
on the correlation length, however, if the system is finite the correlation length saturates 
to $L$, and one has to perform a finite-size scaling. 
In this case the response function in the frequency domain acquires the following scaling form~\cite{Goldenfeld}
\begin{equation}
 \label{scale-transition}
\tilde{\chi}(\omega)=L^{2-\eta}F(\omega L^z)=N^{\frac{2-\eta}{d}}F_1(\omega N^{z/d}),
\end{equation}
where $F(x)$ and $F_1(x)$ are scaling functions,  $z$ is the dynamic exponent, $d$
is the spatial dimension and $\eta$ corresponds to anomalous dimension
exponent. From the previous analysis one sees that
\begin{equation}
 \label{final-form}
\ddot{m}_S(0)=b^2C_r(0)=b^2\int_{-\infty}^{\infty}\frac{d\omega}{2i\pi\omega}\tilde{\chi}(\omega),
\end{equation}
which leads to $\ddot{m}_S(0)=\B N^{\frac{2-\eta}{d}}$,
where the constant $\B=b^2\int_{-\infty}^\infty\frac{dx}{2i\pi x}F_1(x)$. Then we should have, if the general arguments
we have developed before apply, that in the vicinity of a critical point the mean violation fraction
behaves as
\begin{equation}
 \label{mean-phase}
 \langle\nu(\tau,N)\rangle\sim\frac{1}{\tau N^{\frac{2-\eta}{2d}}}.
\end{equation}

The previous result is already different from the general
law given by eq. (\ref{mean-viol-final-fin}) with $\alpha=1$, as should correspond to this case of constant drive,
since $(2-\eta)/d\neq1$ in general.
Furthermore, it could even suggest that in this case
the analysis leading to (\ref{mean-viol-final-fin}) is incorrect because, as said previously,
any finite $N$ is small if the correlation volume of the system is infinite.
In this case the fluctuations are so important, that the asymptotics of $\langle\nu\rangle$ is
not dominated only by the short times.
This fact could also affect the scaling form in the time variable, i.e.,  
our theory breaks down since the typical relaxation time of the system diverges in the thermodynamic limit,
which may render the value of the integral
$\int_0^\infty R(t,N)dt$ infinite.

An interesting case study is, for instance, the mean-field Ising model in two dimensions.
Although it is not a realistic model, in that case one has $(2-\eta)/d=1$ exactly since $\eta=0$. Thus, it
represents a suitable model system to investigate whether the scaling law we have derived here continues to be valid.
This analysis should be, by it self, 
the subject of a separate work. 
\section{Conclusions}
\label{sec:conclusions}
We have introduced and studied the violation fraction as the total fraction of time that the entropy change
during a trajectory in phase space of a mesoscopic system is negative. We have focused on the mean value
of this quantity, which scales, in the large-$\tau$ - large-$N$ limits,
as $\langle\nu(N,\tau)\rangle\sim\big(\tau N^{\frac{1}{1+\alpha}}\big)^{-1}$, with $\alpha>0$
fixed by the time dependence of the external protocol. Our results are robust and independent of the specific
details of the energy landscape and the dynamics, 
as long as the system remains ergodic, microreversible and extensive.
We have also justified that in the vicinity of a critical point, this scaling form
should break down.

Our study is, at this point, still incomplete. First, we have limitated our
derivation to the case of a system prepared in its steady state and connected to
a single heat bath. In this context, we have studied the
Hatano-Sasa dissipation function, which is only a particular form of entropy production.
On the other hand, we have characterized only the first moment of the PDF of the
violation fraction. However, this kind of study is, to the best of our knowledge,
novel. The problem is indeed interesting, since it may help to understand,
for example, how these small fractions of time that a nanomachine spends consuming entropy
affect its efficiency. Furthermore, some interesting questions and open problems
emerge naturally from this study. For example, one may wonder if the violation fraction behaves
non-trivially in the vicinity of a critical point. Does this quantity exhibit singularities in that case?
If in the vicinity of a continuous phase transition the violation fraction grows and correspondingly, the
occurrence of entropy-consuming trajectories is enhanced, then, could we use this fact to improve
the performance of thermal machines? In our opinion, it is worth to try to answer these questions.

Additionally, it would be interesting to define and study the violation fraction
in terms of the entropy production rate as a complement to this study in terms of
the entropy change. An open challenge is to derive a symmetry for the full PDF
of the violation fraction in a more general context, relaxing the necessity of special
initial conditions, allowing the system to be connected to several baths
and considering arbitrary forms of entropy production, and not only the one considered here.
Some of these issues are the subject of a separate paper~\cite{me-next}. 

\begin{acknowledgments}
This work was supported by CNEA, CONICET (PIP11220090100051), and ANPCYT (PICT2011-1537).
We are grateful to the anonymous referees for their useful comments, which helped to highly improve
the quality of this manuscript. RGG would like to specially thank Vivien Lecomte for his valuable
observations during the different stages of this work.
\end{acknowledgments}


\begin{thebibliography}{100}

\bibitem{Seifert-Review}
U. Seifert, Rep. Prog. Phys. {\bf 75}, 126001 (2012)


\bibitem{Evans-Cohen-Morris}
Denis J. Evans, E. G. D. Cohen and G. P. Morriss,  Phys. Rev. Lett. {\bf 71}, 2401 (1993).

\bibitem{Gallavotti-Cohen}
G. Gallavotti and E. G. D. Cohen, Phys. Rev. Lett. {\bf 74}, 2694 (1995).

\bibitem{Kurchan}
J. Kurchan, J. Phys. A: Math. Gen. {\bf 31} 3719 (1998).

\bibitem{Lebowitz}
J. L. Lebowitz and H. Spohn, J. Stat. Phys. {\bf 95} 333 (1999).

\bibitem{Jarzynski}
C. Jarzynski, Phys. Rev. Lett. {\bf 78}, 2690 (1997); C. Jarzynski, Phys. Rev. E {\bf 56}, 5018 (1997).

\bibitem{Crooks}
G. E. Crooks, J. Stat. Phys. {\bf 90}, 1481 (1998); G. E. Crooks, Phys. Rev. E {\bf 61}, 2361 (2000).

\bibitem{Jarzynski1}
V. Y Chernyak, M. Chertkov and C. Jarzynski, J. Stat. Mech. P08001 (2006) 

\bibitem{us1}
R. Garc\'ia-Garc\'ia, D. Dom\'inguez, V. Lecomte, and A. B. Kolton, Phys. Rev. E {\bf 82}, 030104(R) (2010)

\bibitem{us2}
R. Garc\'ia-Garc\'ia, V. Lecomte, A. B. Kolton, and D. Dom\'inguez, J. Stat. Mech. P02009 (2012)


\bibitem{Seifert-non-Markov}
T. Speck and U. Seifert, J. Stat. Mech. L09002 (2007)

\bibitem{Ohkuma-Ohta}
T. Ohkuma and Takao Ohta, J. Stat. Mech. P10010 (2007)

\bibitem{Mai-Dhar}
T. Mai and A. Dhar, Phys. Rev. E {\bf 75}, 061101 (2007)

\bibitem{BiroliFTs}
C. Aron, G. Biroli, and L. F. Cugliandolo, J. Stat. Mech. P11018 (2010) 

\bibitem{Zamponi}
F. Zamponi, F. Bonetto, L. F. Cugliandolo, and J. Kurchan, J. Stat. Mech. P09013 (2005)

\bibitem{Klages}
A. V. Chechkin and R. Klages, J. Stat. Mech. L03002 (2009)

\bibitem{me}
R. Garc\'ia-Garc\'ia, Phys. Rev. E {\bf 86}, 031117 (2012)

\bibitem{Wang}
G. M. Wang, E. M. Sevick, E. Mittag, D. J. Searles, and D. J. Evans, Phys. Rev. Lett. {\bf 89}, 050601 (2002)

\bibitem{Trepagnier}
E. H. Trepagnier, C. Jarzynski, F. Ritort, G. E. Crooks, C. J. Bustamante and J. Liphardt,
PNAS {\bf 101} 15038 (2004)

\bibitem{bustamante}
C. Bustamante, J. Liphardt, and F. Ritort,
Phys. Today, {\bf 58} 43 (2005).

\bibitem{carberry}
D. M. Carberry, M. A. B. Baker, G. M. Wang, E. M. Sevick, and D. J. Evans, J. Opt. A.: Pure Appl. Opt.
{\bf 9}, S204 (2007)

\bibitem{ritort}
F. Ritort,
Adv. Chem. Phys. {\bf 137}, 31 (2008) Ed. Wiley $\&$ Sons. arXiv:0705.0455v1

\bibitem{Gupta}
A. N. Gupta, A. Vincent, K. Neupane, H. Yu, F. Wang, and M. T. Woodside, Nature Phys. {\bf 7}, 631 (2011)

\bibitem{Dornic}
I. Dornic, and C. Godr\`eche, J. Phys. A: Math. Gen. {\bf 31}, 5413 (1998)

\bibitem{Newman}
T.J. Newman, and Z. Toroczkai, Phys. Rev. E {\bf 58}, R2685 (1998)

\bibitem{Hatano-Sasa}
T. Hatano and S.-i. Sasa, Phys. Rev. Lett. {\bf 86}, 3463 (2001)

\bibitem{Parrondo}
J. Prost, J.-F. Joanny and J. M. R. Parrondo, Phys. Rev. Lett. {\bf 103} 090601 (2009)

\bibitem{Goldenfeld}
N. Goldenfeld, {\em Lectures on phase transitions and the renormalization group}
(Perseus Books Publishing, L. L. C., 1992)

\bibitem{me-next}
R. Garc\'ia-Garc\'ia and D. Dom\'inguez,
``Symmetry for the duration of entropy-consuming intervals'', (2013) (unpublished) 
\end{thebibliography}
\end{document}